\documentclass[a4paper,11pt]{article}
\usepackage{pos}
\usepackage{subcaption}
\usepackage{tikz}
\usepackage{verbatim}
\usepackage{bm} 
\usetikzlibrary{positioning,shapes,shadows,arrows}

\newcommand*{\gev}{\textrm{ GeV}}
\newcommand*{\tev}{\textrm{ TeV}}

\title{Studies of Gamma-Ray Shower Reconstruction Using Deep Learning}

\ShortTitle{Energy Reconstruction Using Convolutional Neural Networks}

\author*[a]{Tomas Bylund}
\author[a]{Ga\v{s}per Kukec Mezek}
\author[a]{Mohanraj Senniappan}
\author[a]{Yvonne Becherini}
\author[a,b]{ Michael Punch}
\author[c]{Satyendra Thoudam}
\author[d]{Jean-Pierre Ernenwein}

\affiliation[a]{Linnaeus University, Department of Physics and Electrical Engineering,\\
	35195 V\"axj\"o, Sweden}
\affiliation[b]{Universit\'e de Paris, CNRS, Astroparticule et Cosmologie, F-75013 Paris, France}
\affiliation[c]{ Khalifa University, Department of Physics,\\
	PO Box 127788, Abu Dhabi, United Arab Emirates}
\affiliation[d]{ Aix Marseille Univ, CNRS/IN2P3, CPPM,\\
	Marseille, France.}

\emailAdd{tomas.bylund@lnu.se}
\abstract{
	The Cosmic Multiperspective Event Tracker (CoMET) R\&D project aims to optimize the techniques for the detection of soft-spectrum sources through very-high-energy gamma-ray observations using particle detectors (called ALTO detectors), and atmospheric Cherenkov light collectors (called CLiC detectors). The accurate reconstruction of the energies and maximum depths of gamma-ray events using a surface array only, is an especially challenging problem at low energies, and the focus of the project.
	In this contribution, we leverage Convolutional Neural Networks (CNNs) using the ALTO detectors only, to try to improve reconstruction performance at lower energies ( < 1\tev ) as compared to the SEMLA analysis procedure, which is a more traditional method using manually derived features.
}

\FullConference{37$^{\rm{th}}$ International Cosmic Ray Conference (ICRC 2021)\\
	July 12th -- 23rd, 2021\\
	Online -- Berlin, Germany}

\begin{document}
\maketitle

\section{Introduction}
Recently, surface arrays of particle detectors that are capable of reconstructing very-high-energy (VHE, $ > 200\gev$) gamma-ray induced air showers have become operational \cite{hawc}, bringing with them greatly enhanced fields of view and superior duty cycles compared to Imaging Atmospheric Cherenkov Telescopes (IACT). 
The CoMET project proposes a new design, optimising the particle detector surface arrays for extragalactic targets, meaning that the design should be optimised for high sensitivity at low energies ($< 1$\tev), see \cite{gasper_icrc2021}.
However, the reconstruction of the gamma-ray primary from the lateral distribution of an extensive air shower, as seen by particle detectors, is not an easy feat. Current generation particle detector surface arrays are primarily providing properties, such as the energy, through traditional machine learning methods with hand-constructed features, see for example \cite{hawc}. 

In this contribution, we investigate if convolutional neural networks (CNNs), using only integrated charges from individual ALTO particle detectors, can replace the challenging reconstruction of primary gamma-ray properties.
We present efforts to reconstruct the energy and the depth of shower maximum, $X_{\rm max}$, using a CNN.
The pixel-level information is therefore used directly by a convolutional network to reconstruct the properties of the primary particle.

As a comparison method, we use the SEMLA procedure, see \cite{semla_paper}, a more traditional approach that uses longitudinal fits and derived parameters as input to a staged application of a series of three shallow neural networks. 
In SEMLA, after an initial cut to filter poorly-fitted events, the fit results and other parameters are subjected to further cleaning using a first neural network. 
Then events are tagged as signal-like or background-like using a second network, and then a final network produces a prediction of the event energy. 
For this contribution, using the same input variables and network architecture as for the SEMLA energy evaluation, we performed a SEMLA evaluation of the $X_{\rm max}$ which will be our baseline for the CNN $X_{\rm max}$ evaluation. 

\section{The ALTO detectors in the CoMET array}

\begin{figure*}[t]
	\centering
	\hspace{0.7cm}
	\includegraphics[width=0.38\textwidth] {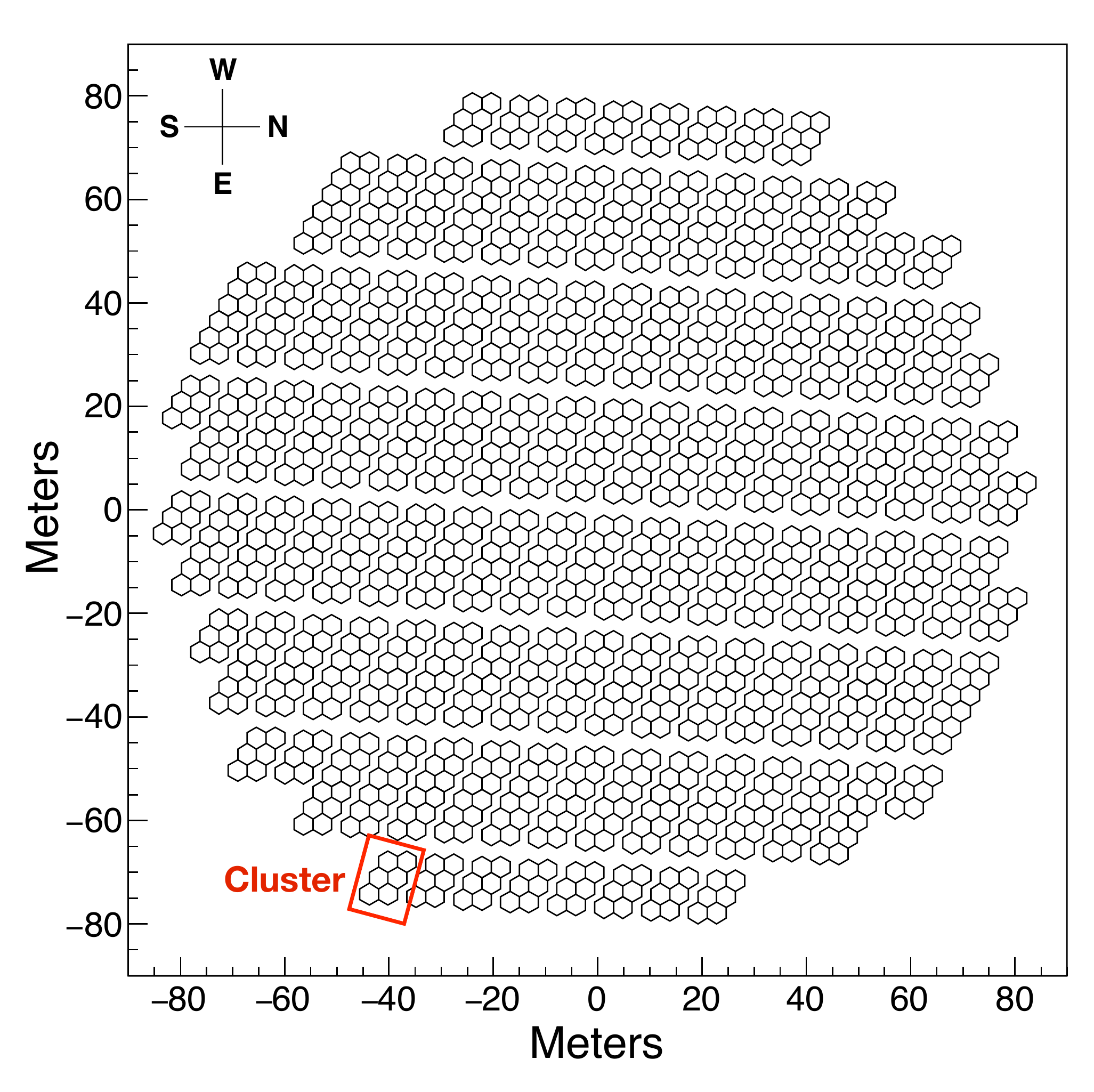}
	\hspace{0.1cm}
	\includegraphics[width=0.48\textwidth] {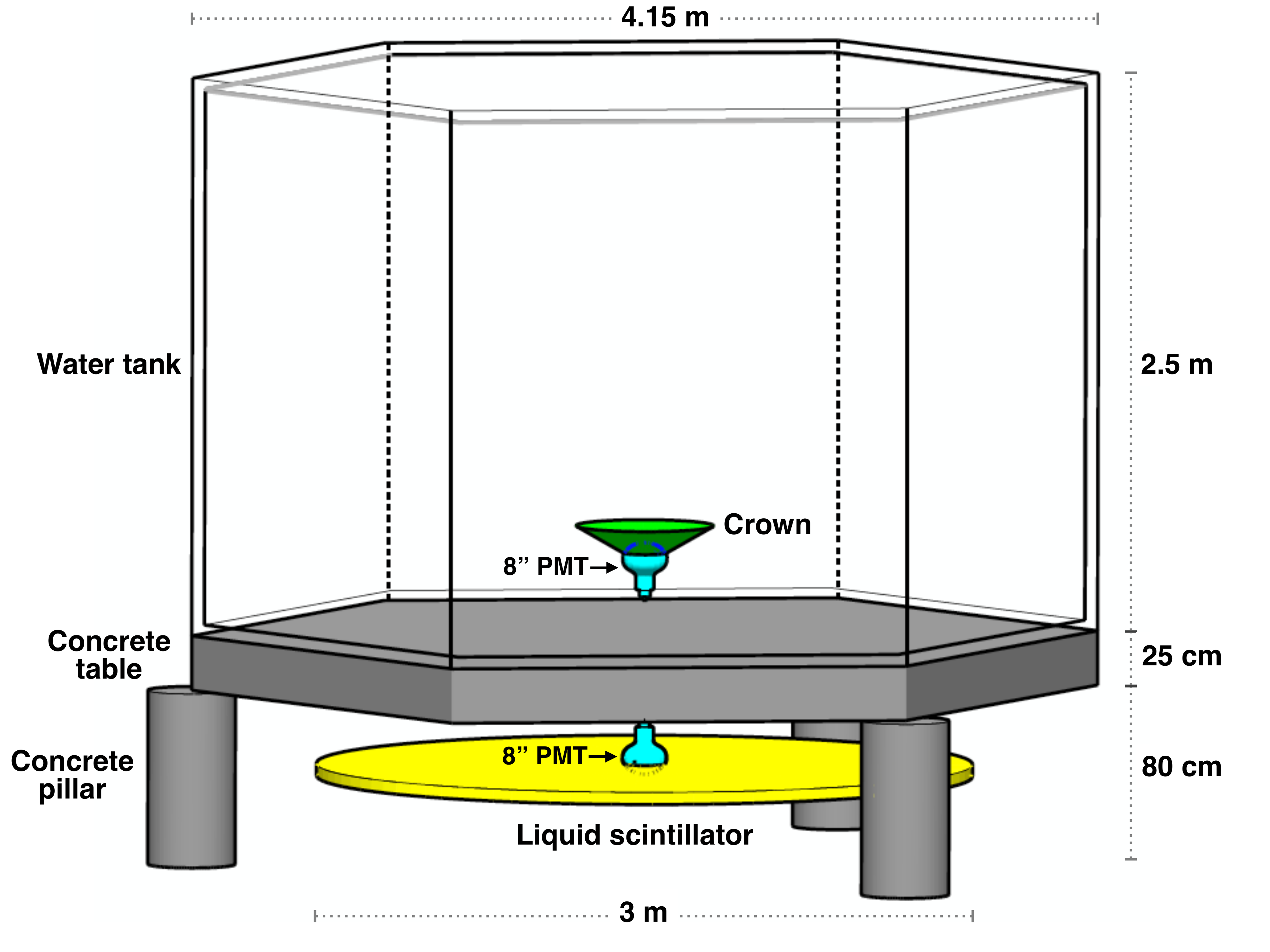}

	\caption{\small{\textbf{ALTO design.} \emph{Left panel:} The layout of the proposed ALTO, note the structure of hexagon clusters separated by smaller spacings. 
			Each \emph{cluster} consists of six ALTO units as highlighted in the red box.
			\emph{Right panel:} The geometry of an ALTO unit with its dimensions. 
			A unit consists of a water Cherenkov detector (WCD), and a liquid scintillator detector (SD) underneath. 
}}

	\label{fig:alto-array}

\end{figure*}

As mentioned earlier, in this contribution we only focus on the ALTO particle detector arrays of the CoMET project.
The ALTO particle detector array consists of 1242 detector units, distributed in a circular area of $\sim 160$ m in diameter, with each unit belonging to a hexagonally packed cluster combining 6 detector units, as can be seen in Figure \ref{fig:alto-array} (left panel).
Each ALTO unit consists of a hexagonal Water Cherenkov Detector (WCD) positioned on a concrete slab, with a cylindrical liquid Scintillator Detector (SD) underneath, see Figure \ref{fig:alto-array} (right panel).
Both the WCD and the SD in one unit are instrumented by an $8^{\prime\prime}$ Hamamatsu Photo Multiplier Tube (PMT) recording the photons emitted by the passage of charged particles.
The WCDs are used for the detection of particles in the air shower, while the SDs are conceived for signal over background discrimination through muon tagging. 
In order to have a low energy threshold for detecting gamma-ray primaries, the CoMET array is intended to be placed at high altitudes ($\sim 5$ km).


\section{The data set}
In this study, we have used gamma-ray events that pass the SEMLA config-Q2 selection cuts \cite{mohan_icrc2021}, aiming to explore improvements of the final reconstruction.
We have reused events from the simulations of the ALTO particle array done for \cite{semla_paper}, so the observatory altitude, magnetic field, primary particle energies and zenith angles have been kept the same. 
In particular, the gamma-rays emulate a point source at $18^{\circ}$ zenith with a spectral index of -2.

The total dataset of gamma-ray events passing the SEMLA stage C cuts amounts to about 164000 events over the energy range 100\gev{} -- 100\tev, and are distributed with a strong central hump peaking at around 1\tev{} (see Figure 12 in \cite{semla_paper}). 
This hump originates from the combination of the effective area energy dependence  and the simulated source spectrum, which represents a fairly generic feature of datasets in VHE gamma-ray astronomy. 
One half of the dataset was used for training and validation, while the other half was held back and only used to estimate the final performance.

\subsection{Event generation}
For the event generation, we perform air-shower simulations using the \textsc{CORSIKA} simulation package (version 7.4387) \cite{corsika}. 
We record the energy of the primary photon being simulated and estimate $X_{\rm max}$, the depth at which the shower contains the maximum number of particles, with a fit to its longitudinal distribution. We consider these values as the Monte Carlo (MC) truth.

The detector response of the ALTO particle array to shower particles was simulated using \textsc{GEANT4} \cite{geant4}, producing arrival times and number of photons at the PMT level. 
These are then processed by our own PMT response simulation to generate electronic waveforms of the signals.
The information saved is the integral of the waveform which is then converted to an estimate of the number of physical photo-electrons, $N_{\rm pe}$, representing the total charge seen in the PMT. 
This number is stored as the main readout of the detector, and is then converted into a rectangular array suitable for the convolutional network using the Oversampling (OV) method. 

\subsubsection{Oversampling}

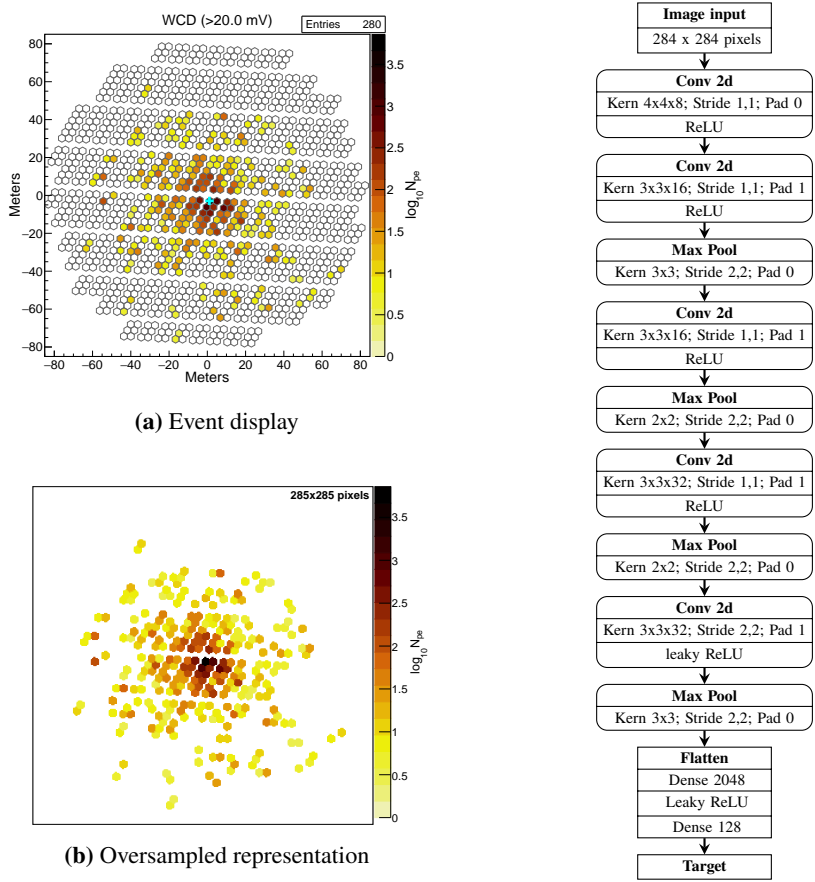
\begin{figure}
\tikzstyle{inputoutput}=[rectangle, draw=black, text centered, anchor=north, text=black, text width=3cm]
\tikzstyle{layer}=[rectangle, draw=black, rounded corners, text centered, anchor=north, text=black, text width=5cm]
\tikzstyle{myarrow}=[->, >=open triangle 90, thick]
\tikzstyle{arrow} = [thick,->,>=stealth]
\tikzstyle{line}=[-, thick]
	\centering
	\hspace{0.9cm}
	\begin{subfigure}{0.37\textwidth}
		\includegraphics[width=\textwidth] {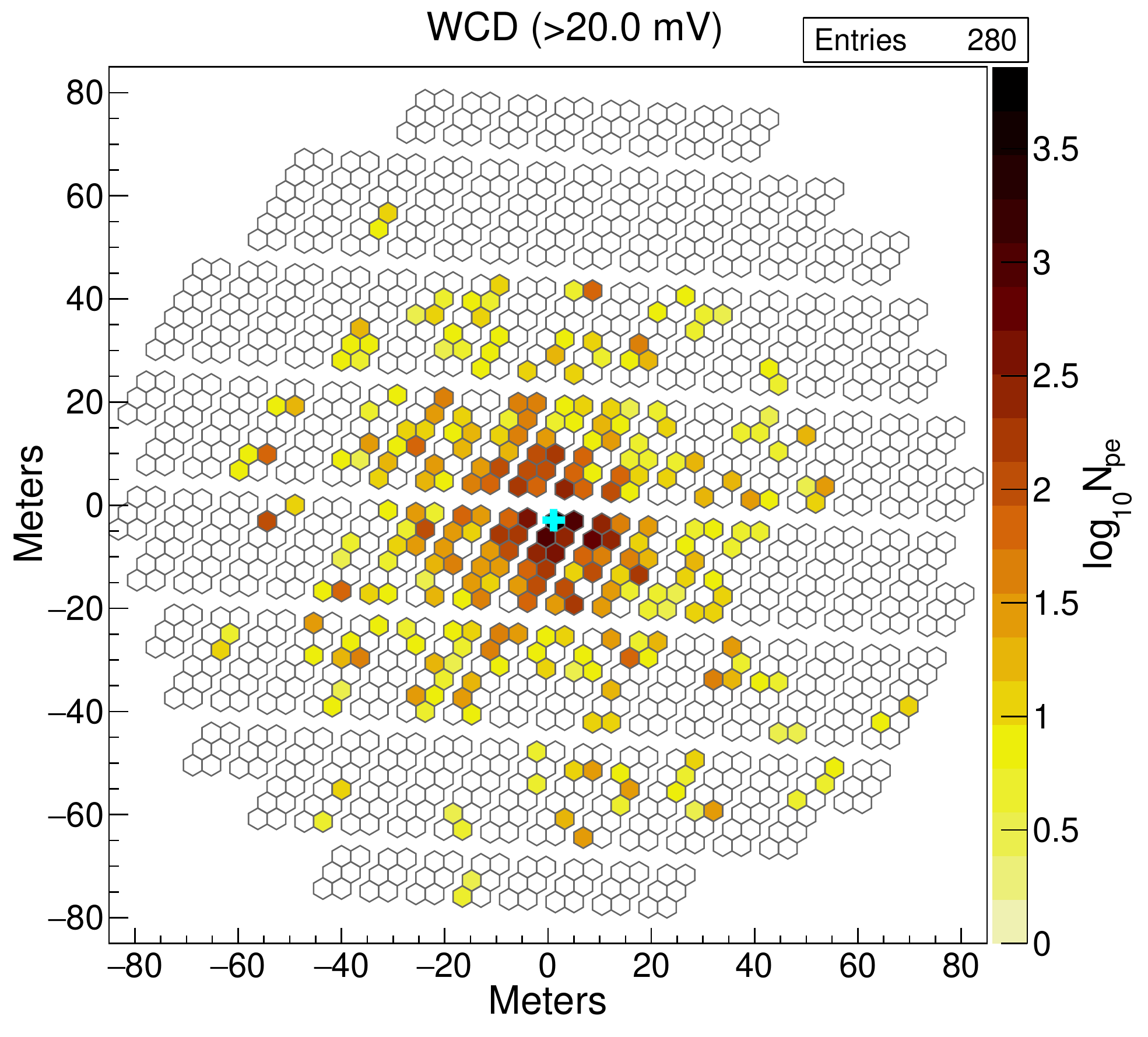}
		\caption{Event display}
		\vspace{0.6cm}
		\hspace{0.28cm}
		\includegraphics[width=0.93\textwidth] {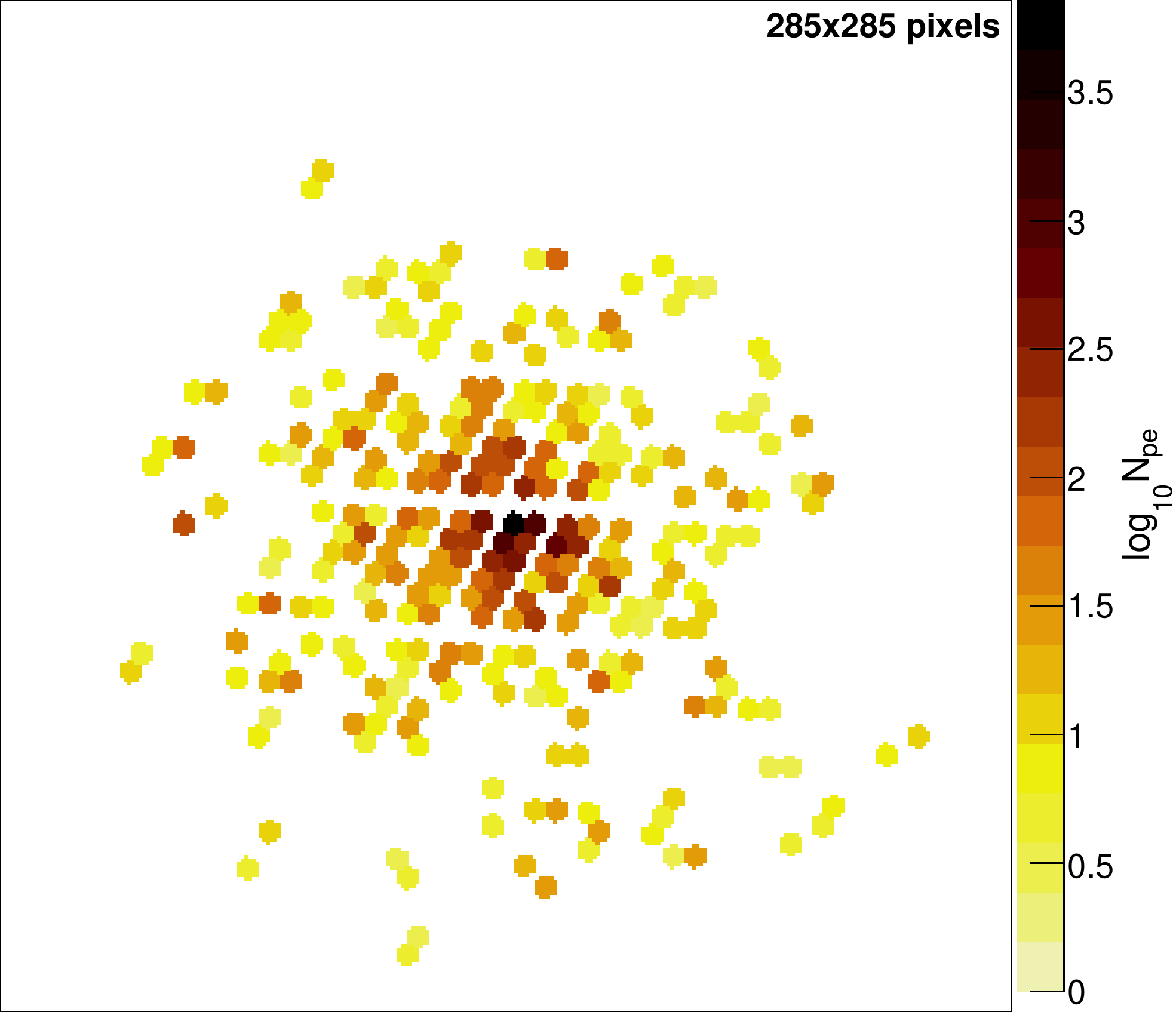} 
		\caption{Oversampled representation} 
	\end{subfigure}
	\begin{subfigure}{0.50\textwidth}
	\hspace{2cm}
	\begin{tikzpicture}[node distance=0.4cm, scale=0.54, every node/.style={transform shape}]
			\node (input) [inputoutput, rectangle split, rectangle split parts=2]
			{
				\textbf{Image input}
				\nodepart{second}{ 284 x 284 pixels}
			};
			\node (conv1) [layer, rectangle split, rectangle split parts=3, below=of input]
			{
				\textbf{Conv 2d}
				\nodepart{second}{Kern 4x4x8; Stride 1,1; Pad 0}
				\nodepart{third}{ReLU}
			};
			\draw[arrow] (input.south) -- (conv1.north);

			\node (conv2) [layer, rectangle split, rectangle split parts=3, below=of conv1]
			{
				\textbf{Conv 2d}
				\nodepart{second}{Kern 3x3x16; Stride 1,1; Pad 1}
				\nodepart{third}{ReLU}
			};
			\draw[arrow] (conv1.south) -- (conv2.north);

			\node (pool1) [layer, rectangle split, rectangle split parts=2, below=of conv2]
			{
				\textbf{Max Pool}
				\nodepart{second}{Kern 3x3; Stride 2,2; Pad 0}
			};
			\draw[arrow] (conv2.south) -- (pool1.north);

			\node (conv3) [layer, rectangle split, rectangle split parts=3, below=of pool1]
			{
				\textbf{Conv 2d}
				\nodepart{second}{Kern 3x3x16; Stride 1,1; Pad 1}
				\nodepart{third}{ReLU}
			};
			\draw[arrow] (pool1.south) -- (conv3.north);

			\node (pool2) [layer, rectangle split, rectangle split parts=2, below=of conv3]
			{
				\textbf{Max Pool}
				\nodepart{second}{Kern 2x2; Stride 2,2; Pad 0}
			};
			\draw[arrow] (conv3.south) -- (pool2.north);

			\node (conv4) [layer, rectangle split, rectangle split parts=3, below=of pool2]
			{
				\textbf{Conv 2d}
				\nodepart{second}{Kern 3x3x32; Stride 1,1; Pad 1}
				\nodepart{third}{ReLU}
			};
			\draw[arrow] (pool2.south) -- (conv4.north);		

			\node (pool3) [layer, rectangle split, rectangle split parts=2, below=of conv4]
			{
				\textbf{Max Pool}
				\nodepart{second}{Kern 2x2; Stride 2,2; Pad 0}
			};
			\draw[arrow] (conv4.south) -- (pool3.north);

			\node (conv5) [layer, rectangle split, rectangle split parts=3, below=of pool3]
			{
				\textbf{Conv 2d}
				\nodepart{second}{Kern 3x3x32; Stride 2,2; Pad 1}
				\nodepart{third}{leaky ReLU}
			};
			\draw[arrow] (pool3.south) -- (conv5.north);

			\node (pool4) [layer, rectangle split, rectangle split parts=2, below=of conv5]
			{
				\textbf{Max Pool}
				\nodepart{second}{Kern 3x3; Stride 2,2; Pad 0}
			};
			\draw[arrow] (conv5.south) -- (pool4.north);

			\node (linear1) [inputoutput, rectangle split, rectangle split parts=4, below=of pool4]
			{
				\textbf{Flatten}
				\nodepart{second}{ Dense 2048 }
				\nodepart{third}{ Leaky ReLU }
				\nodepart{fourth}{ Dense 128 }
			};
			\draw[arrow] (pool4.south) -- (linear1.north);

			\node (output) [inputoutput, below=of linear1]
			{
				\textbf{Target}
			};
			\draw[arrow] (linear1.south) -- (output.north);	
	\end{tikzpicture}
	\end{subfigure}
	\caption{\small{\textbf{Left: Detector representation.} 
			\emph{(a)} Event display for a  1.1\tev{} $\gamma$-ray shower and
			\emph{(b)} a converted image of the same event, using the Oversampling method.
		 }
		\small{\textbf{Right: Network layout.} 
			Layout of the network trained using the oversampled images.
		}}
	\label{fig:networks}
\end{figure}

Following the ideas described in \cite{2019arXiv191209898N}, the complete array is subdivided into numerous square pixels that preserve the spatial distribution of the water tanks. 
Here, we present results using square pixels that correspond to true dimensions of 60 cm to a side, resulting in a 285 by 285 pixel image representing the whole array, see Figure \ref{fig:networks} (a) and (b) for an example of before and after conversion.
Each pixel contains the information on the charge of the tank it falls onto, rescaled as $\log_{10} N_{\rm pe}$. 
Pixels representing areas outside the WCDs are encoded by a default empty value of $-4$.

\section{Convolutional neural networks}

Convolutional neural networks are a type of neural network that (in its initial layers) processes input images using small kernels repeatedly applied across the image. 
The use of the same kernel across the image drastically reduces the number of weights needed, while also making the full network much more robust against random shifts in the location in the image of the object under study \cite{lecun}.
In a typical convolutional network, a series of convolutions and pooling operations are applied, resulting in a progressive decrease of the size of the output image until it is unravelled into a vector of pixels fed into a normal fully connected neural network.
The layout used here is a simple series of gradual downsamplings using unpadded convolutions combined with sharp resizing through \emph{max pooling}, and is illustrated in detail in Figure \ref{fig:networks}.

The network was implemented in PyTorch\footnote{https://pytorch.org/} (1.4.0) using the PyTorch-lightning (1.2.8) framework and trained using the Adam optimiser with default settings and a fixed learning rate of 0.001, in batches of 16, while setting either $\log_{10} E$ or $\log_{10} X_{\rm max}$ as the target variable.



\subsection{Data augmentation}

The CoMET project aims to optimise the detection of soft-spectrum sources, but the impact of the detector effective area results in very few events at lower-bound energies of \mbox{200\gev}.
This means that we have to train our neural networks on a dataset that is imbalanced over our region of interest, and imbalance is known to lead to loss of performance in the classical machine learning setting \cite{branco2016survey}.
We here explore two variants of the ``data pre-processing'' class of solutions to the problem of imbalanced datasets, as opposed to using special loss functions\footnote{We made some attempts in this direction, but typically only achieved slightly worse performance than simply using mean squared error.}.
More specifically, we applied random under- or surplus sampling on the original dataset, to select new samples which we used in the usual train and validation dataset split.

{\bf{Random undersampling.}} Random undersampling means you randomly throw away events until reaching a desired balance between the rare and common events.
We created a finely binned histogram of all the events in the train and validation part of the dataset, and then found the subset of energy bins where each bin contained at least $N_{\rm min}$ number of events. 
The number of bins, the minimum number of events per bin, and the fraction of events used for training were chosen to create a training set consisting of around 5000 events distributed uniformly in a flat energy spectrum. 
Such a small data set results in a faster training, but one limitation of this method is that by requiring a certain minimum number of events per bin, $N_{\rm min}$, we exclude the rare events at the top and bottom end of the spectrum, where $N<N_{\rm min}$.

{\bf{Random surplus sampling.}} Random surplus sampling (also known as oversampling) is the converse idea: instead of discarding events, the sampling is performed by randomly repeating events until the desired target balance is achieved. 
Using a histogram to estimate the density of points $d_i$ inside a narrow bin $i$ with $n_i$ events, we score each event as $(1 - d_i)/n_i$ and convert these scores to weights by dividing them with their total sum.
Finally, using these weights, we augment the population of events by randomly choosing events (with replacement) from the initial sample until our new sample is as flat as desired.
The weighting scheme ensures that rare events in the starting sample are repeated much more frequently in the new sample than the more common ones.
In practice surplus sampling required reusing events to the point that the final sample was 80 \% larger than the starting one to yield a fairly flat spectrum, which means the training is slower.

Finally, we also trained networks using a normal 60\% training, 40\% validation split of the events, so that in total the network was trained using three different datasets: the {\bf{O}}ne {\bf{P}}ass train-validation split, the {\bf{SM}}all undersampled dataset, and the larger {\bf{SU}}rplus sampled dataset.

\section{Results}
\begin{table}[t]
    \centering
    \small
    \begin{tabular}{lccccccc}
        \hline
        \textbf{Configuration} & \textbf{Energy}    & \textbf{Energy }      & $Q_E$  & $\rho_E$ & $\bm{X_{\rm max}}$  &  $\rho_{\bm{X_{\rm max}}}$ & $\bm{X_{\rm max}}$ \\ 
                             & \textbf{training [h]} & \textbf{mean ratio}  &        &          & \textbf{mean ratio} &                            & \textbf{training [h]}\\ 
        
        \hline
            SEMLA & 0.5 & 0.26 & 0.47  & 0.92   & 0.15 &   0.58 & 0.5 \\
            OV OP  & 8.7  & 0.29 & 0.42 & 0.90 & 0.10 & 0.50 & 9.7 \\
            OV SM & 1.6  & 0.25  & 0.43  & 0.90& 0.10 & 0.45 & 1.7 \\
            OV SU & 13.8 & 0.29  & 0.47  & 0.90& 0.10 & 0.48 & 13.0\\
    \end{tabular}
    
      \caption{\small{\textbf{Performance details.} SEMLA is the classical reference implementation using config-Q2 but without the angular separation cut applied in \cite{mohan_icrc2021}. 
			 OV is the oversampled network. 
			 OP refers to a network trained on the One Pass sample, SM on the small sample, and SU on the surplus sampled set.
			 ``Energy training'' or ``$X_{\rm max}$ training'' means the total time taken to train the particular network configuration for 40 epochs for the respective target.
			 $\rho_E$ and $\rho_{\bm{X_{\rm max}}}$ are the Spearman rank correlations between MC truth and predictions.
			 For the definition of the mean ratios, see Section 5. A log ratio of 0.25 corresponds to a relative agreement around 1.8 between the prediction and the MC truth.
			 The SEMLA training time is the end-to-end training time on a new set of events. }}
   \label{table:performance}
\end{table}

As the accuracy of the regression depends strongly on the energy of the event, we adopt the RMS (around zero) of $\log_{10} (a_{\rm reco}/a_{\rm truth})$, with $a$ representing either energy or shower maximum, in bins of true energy as the primary measure.
These RMS curves are summarised by the mean value of the ratio $\log_{10} (a_{\rm reco}/a_{\rm truth})$ expected from each curve, estimated by uniformly sampling energies over the 100\gev\ to 1\tev\ range and converting them to RMS of values before calculating a sample mean, see Table \ref{table:performance} ``mean ratios''.
However, the final evaluation of the quality of an energy estimate is its performance when reconstructing a physical spectrum.
Since implementing a full-forward-folding chain was outside the scope of this work, as a substitute we create a quality score ($Q_{E}$) by averaging the Spearman correlation and the slope of the best fitting line through the transfer matrix for low energy events ($E < 1\tev$), with 1 being the best possible score.
In Table \ref{table:performance} we also give the Spearman rank correlation $\rho_a$ for the entire test sample.

\begin{figure*}[t]
	\centering
	\begin{subfigure}{0.47\textwidth}
		\includegraphics[width=\textwidth] {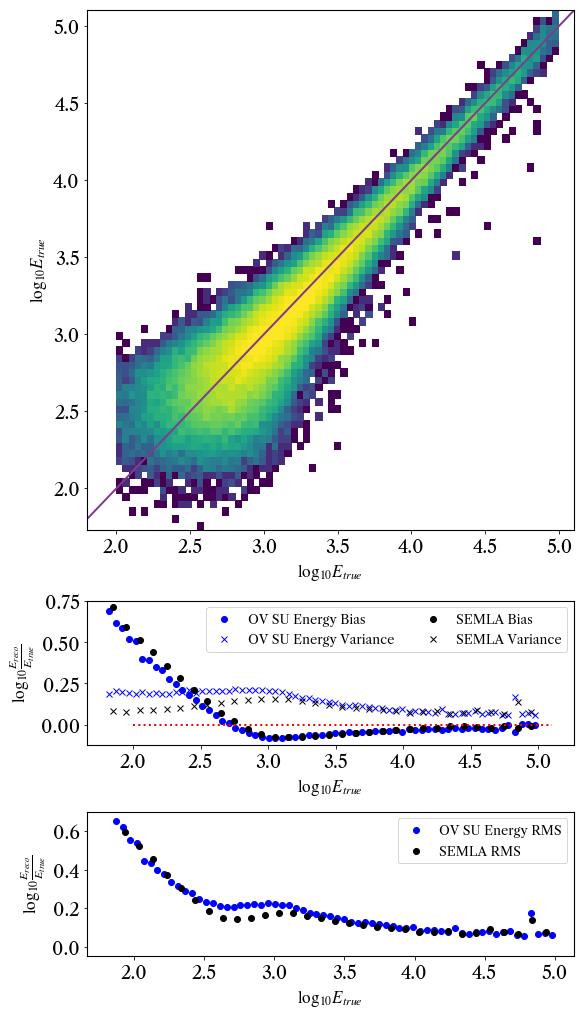}
	\end{subfigure}
	\begin{subfigure}{0.48\textwidth}
		\includegraphics[width=\textwidth] {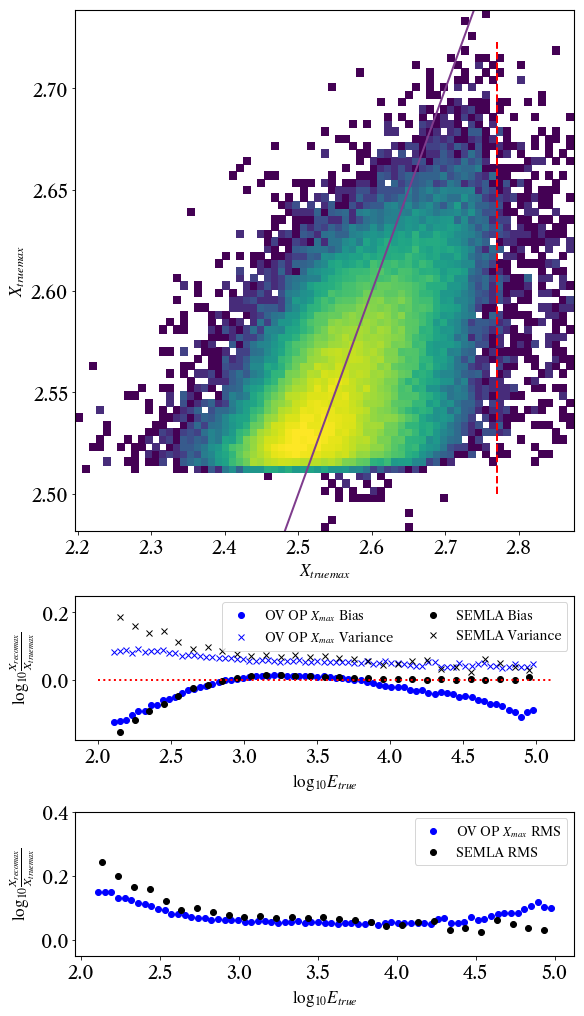}
	\end{subfigure}

	\caption{\small{\textbf{Analysis performance.} \emph{Left column:} the transfer matrix, bias, variance and resolution of $\log_{10} E_{\rm reco}/E_{\rm true}$  for the Oversampled network with surplus sampled data. 
		\emph{Right column:} the transfer matrix, bias, variance and resolution of $\log  X_{\rm recomax}/X_{\rm truemax}$  for the Oversampled network with One Pass data. 
		 Blue markers are this work, black points is the SEMLA config-Q2 w/o angular separation cuts.
		 The purple line shows perfect correlation, and the red dashed line shows the depth of the ALTO detector.
	}}
			\label{fig:performance}
			\vspace{-0.1cm}
\end{figure*}

Figure \ref{fig:performance} shows a comparison of the best performing model in terms of resolution and quality, along with the SEMLA result for both regression targets. 
The SEMLA results shown are obtained with a ``One Pass'' configuration, as the spectrum resampling procedure did not improve the performance.
In Table \ref{table:performance} we provide only rough training times in the form of total time for 40 epochs of training to give a sense of the time required. 

{\bf{Energy regression.}} In the case of energy regression, the traditional SEMLA analysis proves hard to beat, with the CNNs unable to match its overall performance (see bottom left panel, Figure \ref{fig:performance}), though none of the four methods manage better than 0.25 mean log ratio in the low-energy region, see Table \ref{table:performance}. 
The $Q_E$ scores in Table \ref{table:performance} indicate that the oversampled network benefits somewhat from training on the surplus sampled dataset (OV SU) but without any clear benefit in the mean ratios or overall correlations $\rho_E$, thus suggesting that the other methods have achieved better scores by introducing unphysical non-linearity in the transfer matrix.
  
{\bf{$\bm{X_{\rm max}}$ regression.}} The situation is however reversed when we examine the results for $X_{\rm max}$, where the CNNs consistently beat the traditional machine learning approach. 
These predictions are all however of modest power, with fairly low-quality score and reaching a Spearman correlation of 0.5 between the true and the reconstructed value, see Table \ref{table:performance}.
Since more sophisticated approaches -- using the timing information in the signal traces from WCDs, not only the total integrated charge -- have reached correlations of 0.8 \cite{auger}, there is hope that the limited performance presented here can be substantially improved.

From the transfer matrix in Figure \ref{fig:performance} we observe that high-altitude events are correctly reconstructed far away from the detector location, and this information might be useful in a future version of the ALTO reconstruction and analysis chain with the aim of improving the performance at low energies.

\section{Conclusions}

We have presented attempts to reconstruct the energy of VHE gamma-rays, and $X_{\rm max}$ of the resulting air showers, using only the total charge seen in the ALTO particle detector surface array with convolutional neural networks.
The CNNs performed comparable to a classical feature-based machine learning approach when evaluating the energy, which is in line with previous findings using IACT data stating that the energy reconstruction of existing methods can be hard to beat \cite{dan}.
This result could be due to the good understanding of the physical connection between shower properties and initial energy.
In contrast, CNNs performed uniformly better than the classical feature-based machine learning when predicting $X_{\rm max}$, meaning that the convolutional networks are capable of extracting physical features from the detector array data, where no obvious correlation between the observables and $X_{\rm max}$ have be found manually.

\section{Acknowledgements}

\url{https://alto-gamma-ray-observatory.org/acknowledgements/}

\bibliographystyle{JHEP}
\bibliography{ICRC2021}

\providecommand{\href}[2]{#2}\begingroup\raggedright\begin{thebibliography}{10}

\bibitem{hawc}
A.U.~{Abeysekara}, A.~{Albert}, R.~{Alfaro}, C.~{Alvarez}, J.D.~{{\'A}lvarez},
  J.R.A.~{Camacho} et~al., \emph{{Measurement of the Crab Nebula Spectrum Past
  100 TeV with HAWC}},
  \href{https://doi.org/10.3847/1538-4357/ab2f7d}{\emph{The Astrophysical
  Journal} {\bfseries 881} (2019) 134}
  [\href{https://arxiv.org/abs/1905.12518}{{\ttfamily 1905.12518}}].

\bibitem{gasper_icrc2021}
G.~{Kukec Mezek}, M.~{Senniappan}, Y.~{Becherini}, M.~{Punch}, S.~{Thoudam},
  T.~{Bylund} et~al., \emph{{The COMET multiperspective event tracker for wide
  field-of-view gamma-ray astronomy}},
  \href{https://doi.org/10.22323/1.358.0270}{\emph{PoS} {\bfseries ICRC2021}
  (2021) at this conference.}

\bibitem{semla_paper}
M.~{Senniappan}, Y.~{Becherini}, M.~{Punch}, S.~{Thoudam}, T.~{Bylund},
  G.~{Kukec Mezek} et~al., \emph{{Signal extraction in atmospheric shower
  arrays designed for $\rm 200\,GeV-50\,TeV$ $\gamma$-ray astronomy}},
  {\emph{arXiv e-prints} (2021) arXiv:2105.06728}
  [\href{https://arxiv.org/abs/2105.06728}{{\ttfamily 2105.06728}}].

\bibitem{mohan_icrc2021}
M.~{Senniappan}, Y.~{Becherini}, M.~{Punch}, S.~{Thoudam}, T.~{Bylund},
  G.~{Kukec Mezek} et~al., \emph{{Expected performance of the ALTO particle
  detector array designed for 200 GeV - 50 TeV gamma-ray astronomy}},
  \href{https://doi.org/10.22323/1.358.0270}{\emph{PoS} {\bfseries ICRC2021}
  (2021) at this conference.}

\bibitem{corsika}
D.~{Heck}, J.~{Knapp}, J.N.~{Capdevielle}, G.~{Schatz} and T.~{Thouw},
  \emph{{CORSIKA: a Monte Carlo code to simulate extensive air showers.}}
  (1998).

\bibitem{geant4}
{\scshape GEANT4} collaboration, \emph{{GEANT4--a simulation toolkit}},
  \href{https://doi.org/10.1016/S0168-9002(03)01368-8}{\emph{Nucl. Instrum.
  Meth. A} {\bfseries 506} (2003) 250}.

\bibitem{2019arXiv191209898N}
D.~{Nieto}, A.~{Brill}, Q.~{Feng}, M.~{Jacquemont}, B.~{Kim}, T.~{Miener}
  et~al., \emph{{Studying deep convolutional neural networks with hexagonal
  lattices for imaging atmospheric Cherenkov telescope event reconstruction}},
  {\emph{arXiv e-prints} (2019) arXiv:1912.09898}
  [\href{https://arxiv.org/abs/1912.09898}{{\ttfamily 1912.09898}}].

\bibitem{lecun}
Y.~LeCun, L.~Bottou, Y.~Bengio and P.~Haffner, \emph{Gradient-based learning
  applied to document recognition}, {\emph{Proceedings of the IEEE} {\bfseries
  86} (1998) 2278}.

\bibitem{branco2016survey}
P.~Branco, L.~Torgo and R.P.~Ribeiro, \emph{A survey of predictive modeling on
  imbalanced domains}, {\emph{ACM Computing Surveys (CSUR)} {\bfseries 49}
  (2016) 1}.

\bibitem{auger}
J.~Glombitza, \emph{{Air-Shower Reconstruction at the Pierre Auger Observatory
  based on Deep Learning}},
  \href{https://doi.org/10.22323/1.358.0270}{\emph{PoS} {\bfseries ICRC2019}
  (2019) 270}.

\bibitem{dan}
R.D.~{Parsons} and S.~{Ohm}, \emph{{Background rejection in atmospheric
  Cherenkov telescopes using recurrent convolutional neural networks}},
  \href{https://doi.org/10.1140/epjc/s10052-020-7953-3}{\emph{European Physical
  Journal C} {\bfseries 80} (2020) 363}
  [\href{https://arxiv.org/abs/1910.09435}{{\ttfamily 1910.09435}}].

\end{thebibliography}\endgroup

\end{document}